# Predicting the Expansion of Concrete Exposed to Sulfate Attack with a Regression Model Based on a Performance Classification

Xiangru Jian[1], Paulo J.M Monteiro[2] and Kimberly E. Kurtis[3]

[1]College of Civil Engineering, Tongji University, Shanghai 200092, China

[2]Department of Civil and Environmental Engineering at the University of California at Berkeley

[3]School of Civil and Environmental Engineering at the Georgia Institute of Technology

**Abstract**

This paper mainly describes the development of a new type of regression model to predict the long-term expansion of concrete subjected to a sulfate-rich environment. The experimental data originated from a long-term (40+ years), nonaccelerated test program performed by the U.S. Bureau of Reclamation (USBR). Expansion data of specimens composed of 54 different mixtures were measured periodically throughout the entire test program. In this analysis, the mixtures were first classified into three groups using K-means clustering based on their expansion patterns. Within each group, the expansion rate was predicted as an exclusive regression function of the water-cement ratio (W/C), tricalcium aluminate ($C_3A$) content of cement, cement content of cement or time of expansion. Then, a support vector machine (SVM) was employed to determine the classification criteria by relying on the characteristics of the mixture proportions rather than the experimental performance, thereby enabling the model to offer predictions for new mixtures without test data. An analysis of the model indicated that concrete specimens with different mixture proportions, especially with different W/C values or $C_3A$ contents, are unlikely to share identical expansion patterns and should be considered and predicted separately.

## 1 Introduction

Civil engineers design most structures for 50–100 years of safe performance with minimal maintenance. However, several chemicals are likely to pose a major threat to the long-term durability of concrete and could lead to the nonmechanical failure of the structure over time, especially when the structure is exposed to severe and aggressive environments. Among all of the failure mechanisms associated with chemicals, the sulfate attack failure mechanism is the most significant for structures built upon or surrounded by sulfate-rich environments such as soils, groundwater, seawater, decaying organic matter, and industrial effluent. Sulfate ions from outside sources can react with hydrated Portland cement and potentially cause volumetric expansion; as a consequence, the concrete will crack, and the degradation of the concrete will accelerate as cracks propagate and spalling occurs. Ultimately, the concrete will exhibit an increased permeability in addition to a progressive loss of both mass and strength. Nevertheless, the demand for construction in harsh environments continues to increase for economic reasons. Therefore, investigating methods of preventing and alleviating the deterioration, degradation and even severe failure of concrete may constitute the most important components of concrete durability research[1-4].

However, in the absence of a comprehensive database describing the long-term performance of concrete that is exposed to sulfate environments typical of field conditions, selecting the most suitable chemical composition and mixing proportions of concrete materials to ensure the resulting concrete has sufficient resistance to sulfate attack can be difficult. Thus, paralleling the conclusions of Cohen and Mather[5], to compensate for the scarcity of information due to limited experimental data, there is an urgent need to develop more reliable models that can predict the performance of concrete materials exposed to sulfate ions over a service life spanning from several hundreds to several thousands of years.

A nonaccelerated laboratory test program was initiated by the U.S. Bureau of Reclamation (USBR) in the early 1940s to determine the influences of a variety of concrete mixture parameters on sulfate resistance. Under this comprehensive experimental program, which spanned 40 years (it was completed in 1991), concrete samples were immersed in sulfate solutions, and the expansion of the concrete was measured over time. Subsequently, many methodologies have been employed to analyze this comprehensive dataset and develop a new predictive model. Kurtis et al.[6] used panel data methods to identify some of the key variables that influence the expansion of concrete. According to the result , concrete expansion increases over time with increasing water-cement ratio (W/C) and tricalcium aluminate ($C_3A$) content of cement, especially when $C_3A$ content is over 8%. Corr et al.[7] conducted a reliability analysis based on statistical equations for concrete mixtures with $C_3A$ contents of cement under 8%; the results showed that W/C is the most dominant parameter of the parameters included in the reliability model, which included exposure time and $C_3A$ content. The influence of W/C is roughly twice as significant as exposure time and ten times larger than the $C_3A$ content..

Monteiro and Kurtis[8] correlated the failure time of testing samples to their W/C, cement composition, and replacement percentage of cement with fly ash. The analysis indicated that failure did not occur for specimens with a W/C lower than 0.45 and a $C_3A$ content of cement lower than 8%, even when exposed to a sulfate-rich environment over the 40 years . Monteiro et al.[9] proposed scaling laws for the expansion pattern of concrete exposed to a sodium sulfate solution. The results showed that the expansion of concrete samples with a W/C over 0.5 followed a definite scaling law after an initiation period. In addition, their analysis demonstrated that the cement composition most influences the scaling exponent of the model rather than the original W/C. The initiation time depends on both the W/C and the cement composition. Haj-Ali et al.[10] trained an artificial neural network (ANN) model to predict the expansion of concrete with inputs of exposure time, W/C and $C_3A$ content of cement. The model proved effective and accurate at predicting concrete expansion over a span of 40 years, even though the amount of useful data from USBR for ANN training was quite limited.

Admittedly, the abovementioned studies developed useful models for predicting concrete expansion under sulfate attack using different methodologies. However, these studies failed to consider that mixtures consisting of different components and varying W/C values might have completely different expansion patterns and expansion parameters when exposed to sulfate-rich environments. Without delineating clear boundaries between mixtures that do not share the same expansion patterns and parameters to analyze them separately, it will be quite difficult to offer practical guidelines for civil engineers to choose a mixture with adequate sulfate resistance.

Therefore, the objective of this research is to develop several regression models based on mixture proportions to predict the expansion of concrete in sulfate-rich environments. The K-means clustering algorithm is applied to classify mixtures into groups by examining the patterns of expansion associated with the mixtures. Then, every group of mixtures will obtain a unique predictive linear regression model with the most suitable and dominant parameters. As determined by principal component analysis (PCA). Subsequently, because there is no direct expression of the classification criterion based only on the properties of the mixture proportion other than expansion data, which can be obtained by K-means clustering, clear boundaries between the groups will be demarcated based on mixture components and proportions using the support vector machine (SVM) algorithm, thereby enabling us to predict the expansion data of a certain mixture. All of the developed models are based upon the results from the abovementioned nonaccelerated tests performed by the USBR in Denver, Colorado, during which the expansion characteristics of concrete cylinders exposed to severe sulfate conditions were measured over a period exceeding 40 years. These data are important because they represent the only comprehensive database for concrete materials exposed to the typical field concentrations of sulfates over such a long period. In addition, the sulfate resistance has not yet been monitored over such an extended timeframe with concrete specimens prepared from such a wide variety of mixtures.

**2 Mathematical Formulation**

This section addresses the definitions of the algorithms, beginning with the K-means clustering algorithm, that are relevant to this research.

**2.1 K-means clustering**

In general, K-means clustering aims to partition ***n*** observations into ***k*** clusters, with each observation belonging to the cluster with the nearest mean and serving as a prototype of the cluster. This method results in a partitioning of the data space into Voronoi cells[11]. Specifically, the K-means clustering process can be divided into four steps as follows[12].

Initialization. Set the K-means $\{m^{(k)}\}$ to random values.

Assignment. Each data point $x^{(n)}$ is assigned to the nearest mean. We denote our estimate for the cluster $k^{(n)}$ to which the point $x^{(n)}$ belongs as $\hat{k}^{(n)}$.

$$\hat{k}^{(n)} = \arg\min_{k}\{d(m^{(k)}, x^{(n)})\} \quad (1)$$

If two or more means are exactly the same distance from a data point, $\hat{k}^{(n)}$ will be assigned the smallest value of $\{k\}$. Additionally, in this step, we need to introduce the indicator variable $r_k^{(n)}$, which is set to one if the mean k value is the closest mean to the data point $x^{(n)}$, which implies that the mean is responsible for the specific data point; otherwise, $r_k^{(n)}$ is zero.

$$r_k^{(n)} = \begin{cases} 1 & if \quad \hat{k}^{(n)} = k \\ 0 & if \quad \hat{k}^{(n)} \neq k \end{cases} \quad (2)$$

Update. The means are adjusted to match the sample means of the data points for which they are responsible, as follows:

$$m^{(k)} = \frac{\sum_n r_k^{(n)} x^{(n)}}{R^{(k)}} \quad (3)$$

where R(k) is the total responsibility of the mean k:

$$R^{(k)} = \sum_n r_k^{(n)} \quad (4)$$

If $R^{(k)} = 0$, then we leave the mean $m^{(k)}$ where it is.

We repeat the assignment step and update step until the assignments do not change. Then, we obtain all K-means and k clusters that contain the data points we have.

**2.2 Principal component analysis**

PCA[13] is a "statistical procedure that uses an orthogonal transformation to convert a set of observations of possibly correlated variables into a set of values of linearly uncorrelated variables called principal components"

Specifically, PCA finds a few new uncorrelated parameters that are linear combinations of the original correlated parameters to account for most of the variation in the dataset[14]. In this way, PCA transforms the original coordinate system into a new coordinate system, where the projection of the dataset on the first coordinate (i.e., the first principal component) has the greatest variance among all projections, the second coordinate has the second greatest variance, and so on. The obtained principle components together establish an uncorrelated orthogonal basis set. Theoretically, the number of principal components is equal to the number of parameters and observations minus one; however, in most cases, the first few principal components are capable of explaining most of the total variation.

Consider a data matrix X with a column-wise zero empirical mean in which each of the n rows represents a different sample and each of the p columns represents a particular kind of feature. To determine the new coordinates, we need to define several p-dimensional unit vectors, called loading vectors and denoted by $w_{(k)} = (w_1, \dots, w_p)_{(k)}$, with which we can transform the old coordinate system into a new coordinate system. Because the loading vectors are unit vectors, the square modulus of the vector $Xw_{(k)}$, which is the product of the data matrix $X$ multiplied by a certain loading vector $w_{(k)}$, shows exactly the maximum possible variance of the data in the matrix $X$ under the direction given by the loading vector $w_{(k)}$.

To maximize variance, the first loading vector $w_{(k)}$ must lead to the maximum possible variance.

$$w_{(1)} = \underset{\|w\|=1}{arg\ max} \{\|Xw\|^2\} \quad (5)$$

The $k$th component can be found based on the same equation using the new data matrix $\widehat{X_k}$ by subtracting the first $k-1$ principal components from $X$.

$$\widehat{X_k} = X - \sum_{i=1}^{k-1} Xw_{(i)} w_{(i)}^T \quad (6)$$

$$w_{(k)} = \arg\max_{\|w\|=1} \{\|\widehat{X_k}w\|^2\} \quad (7)$$

In this way, all $p$ loading vectors can be derived, and thus, all principal components can be obtained.

## 2.3 Linear regression with the ordinary least squares method

In statistics, linear regression is a linear approach used to model the relationship between a scalar response and one or more explanatory variables. When there is more than one explanatory variable, this process is called multiple linear regression.

To be more explicit, given a dataset $\{y_i, x_{i1}, \dots, x_{ip}\}_{i=1}^n$ of n statistical units, a linear regression model assumes that the relationship between the dependent variable y and the p-vector composed of regressors x is linear. This relationship is modeled through a disturbance term or error variable ε, which represents an unobserved random variable that adds a deviation between the dependent variable and regressors. Thus, the model can be written in matrix notation as follows:

$$y = X\beta + \varepsilon \quad (8)$$

where:

$$y = \begin{pmatrix} y_1 \\ y_2 \\ \vdots \\ y_n \end{pmatrix} \quad X = \begin{pmatrix} x_1^T \\ x_2^T \\ \vdots \\ x_n^T \end{pmatrix} = \begin{pmatrix} 1 & x_{11} & \cdots & x_{1p} \\ 1 & x_{21} & \cdots & x_{2p} \\ \vdots & \vdots & \ddots & \vdots \\ 1 & x_{n1} & \cdots & x_{np} \end{pmatrix}$$

$$\beta = \begin{pmatrix} \beta_0 \\ \beta_1 \\ \vdots \\ \beta_p \end{pmatrix} \quad \varepsilon = \begin{pmatrix} \varepsilon_1 \\ \varepsilon_2 \\ \vdots \\ \varepsilon_n \end{pmatrix}$$

In the above form, y is a vector of observed values $y_i (i = 1, \dots, n)$ that represent the variable, which is called the regressand, X can be regarded as a matrix of row vectors $x_i$, which are known as the regressors, and thus, X can be called the regressor matrix, $\beta$ is a (p+1)-dimensional parameter vector whose elements are known as regression coefficients, and $\varepsilon$ is a vector of the error variable $\varepsilon_i$. Of these parameters, the parameter vector $\beta$ must be estimated in the linear regression model. Several estimation methods are used very frequently for this purpose. In our model, the ordinary least squares (OLS) method is chosen due to its simplicity and high efficiency.

Suppose $b$ is a possible value for the parameter vector $\beta$. The quantity $y_i - x_i^T b$, which is the residual for the i-th observation, measures the vertical distance between the data point $(x_i, y_i)$ and the hyperplane $y = x_i^T b$ and thus assesses the degree of fit between the actual data and the model. The residual sum of squares (RSS)[15] is a measure of the overall model fit:

$$S(b) = (y - Xb)^T(y - Xb) \quad (9)$$

The value of $b$ that minimizes $S(b)$ is called the OLS estimator of $\beta$. The function $S(b)$ is quadratic in $b$ with a positive definite Hessian; therefore, this function possesses a unique global minimum at $b = \hat{\beta}$, which can be given by the following explicit formula[15]:

$$\hat{\beta} = (X^T X)^{-1} X^T y \quad (10)$$

After we have estimated $\beta$, the fitted values (or predicted values) from the regression will

be as follows:

$$\hat{y} = X\hat{\beta} \quad (11)$$

It is common to assess the goodness of fit of the OLS regression by calculating how much the initial variation in the sample can be reduced by regressing onto $X$. The coefficient of determination $R^2$ is defined as a ratio of the variance that can be explained by the regression results to the total variance of the dependent variable y (where $\bar{y}$ is the average of $y$)[15]:

$$R^2 = \frac{\sum(\hat{y}_i - \bar{y})^2}{\sum(y_i - \bar{y})^2} \quad (12)$$

Here, $R^2$ will always be a number between 0 and 1, where values close to 1 indicate a good degree of fit.

**2.4 Support vector machine**

Support vector machines (SVMs)[16] are "supervised learning models with associated learning algorithms that analyze data used for classification and regression analysis."(Cortes et al,1995) With a set of training examples, an SVM training algorithm constructs a non-probabilistic binary linear classifier that assigns new examples to one of two categories. Specifically, a general SVM helps construct a hyperplane that can be used for classification and regression in a high-dimensional space. The hyperplane is built to maximize the distance to the nearest data point of both classes (so-called functional margin) to provides the best separation, as a larger margin generally results in a lower generalization error of the classifier.

The detailed process is as follows. Suppose we have a set of training data, where $x_n$ is a multivariate set of $N$ observations with observed response values $y_n$. The goal of the SVM algorithm is to find the linear function with the largest margin to the nearest training data point (which forms the support vector):

$$f(x) = x'\beta + b \quad (13)$$

This task is equivalent to finding the function $f(x)$ with the minimum norm $(\beta'\beta)$, which is formulated as a convex optimization problem to minimize $J(\beta)$ as follows:

$$J(\beta) = \frac{1}{2}\beta'\beta \quad (14)$$

where all residuals must have a value less than a given allowable variation ε, as shown in the equation below:

$$\forall n: \left|y_n - \left(x'_n\beta + b\right)\right| < \varepsilon \quad (15)$$

It is possible that no such function *f(x)* exists to satisfy these constraints for all points, especially in some complicated scenarios with large errors in the training data. To address otherwise infeasible constraints, we introduce two slack variables, namely, $\xi_n$ and $\xi_n^*$, for each point. This approach is called "soft margin" in SVM classification, as regression errors are allowed to exist up to the value of the slack variables, $\xi_n$ and $\xi_n^*$.

Introducing the slack variables into $J(\beta)$ leads to a new objective function, known as the primal formula, as shown below:

$$J(\beta) = \frac{1}{2}\beta'\beta + C \sum_{n=1}^{N}(\xi_n + \xi_n^*) \quad (16)$$

which is subject to the following constraints:

$$\forall n: y_n - (x_n'\beta + b) < \varepsilon + \xi_n \quad (17)$$

$$\forall n: (x_n'\beta + b) - y_n < \varepsilon + \xi_n^* \quad (18)$$

$$\forall n: \xi_n^* > 0 \quad (19)$$

$$\forall n: \xi_n > 0 \quad (20)$$

The constant $C$ in the objective function is called the box constraint, and it is a positive coefficient that controls the penalty imposed on outliers whose error surpasses the margin $\varepsilon$ and thus helps to avoid overfitting. This value determines the trade-off between the magnitude of the margin between $f(x)$ and the support vector and the extent to which deviations larger than $\varepsilon$ are tolerated.

The optimization problem of the primal formulation is computationally simpler to solve in its Lagrange dual formulation, which is not mentioned herein because it is more complicated and constitutes only an alternative computational method.

After optimizing the objective formula, we can obtain the coefficient $\beta$ and therefore the classifier function $f(x)$.

### 3 Results and Discussion
### 3.1 Classification using K-means clustering

As mentioned above, different mixtures of concrete materials may not share the same magnitude or pattern of expansion under sulfate attack [6]. Accordingly, it is highly necessary to classify concrete materials of different mixing proportions into several groups that share similar expansion patterns before performing linear regression. In this study, we consider 54 different mixtures instead of specimens because mixtures and specimens exhibit a one-to-one corresponding relationship, and the experimental data of specimens are just a reflection of the properties of the mixtures. Moreover, because the expansion pattern needs to be classified, the properties of the expansion curves will be more helpful during the classification than the mixture parameters, that is, the time at which the concrete fails under sulfate attack and the expansion speed at that time constitute the most important criteria. According to relevant work, the failure of concrete under sulfate attack occurs when the expansion rate surpasses 0.5%. Therefore, the coordinates of the failure point and its slope are used as the criteria in the K-means clustering algorithm to divide the data into k groups. This selection step is critical because it determines the total number of groups and thus the quantity of classifications. According to Monteiro et al.[6], concrete mixtures are divided into two groups based on two constituents, namely, W/C and $C_3A$, which represent the properties of the mixture. However, our research focuses more on the direct behavior of concrete, namely, the expansion curve, under sulfate attack, and thus, the results are considered to be more accurate. Therefore, based on the expansion curves of the 54 mixtures, three groups with clear boundaries between them can be found.

From the results illustrated in Fig. 1, all three groups exhibit completely different speeds of expansion. A relatively high and nonlinear expansion rate is observed for the first group, which has a very early failure point, and thus, we call it a group with a high speed and nonlinear curve (group HN). he second group (in the center of Fig. 1) has a service life of approximately 20 years before failure and exhibits a moderate speed and linear expansion; thus, this group of mixtures possesses a moderate speed and a linear curve (group ML). The third group failed near the end of the testing period because it was characterized by a very low expansion rate and a linear expansion curve; consequently, we label this group with a low speed and a linear curve (group LL). Hereafter, all work (including regression) will be conducted for these three groups.

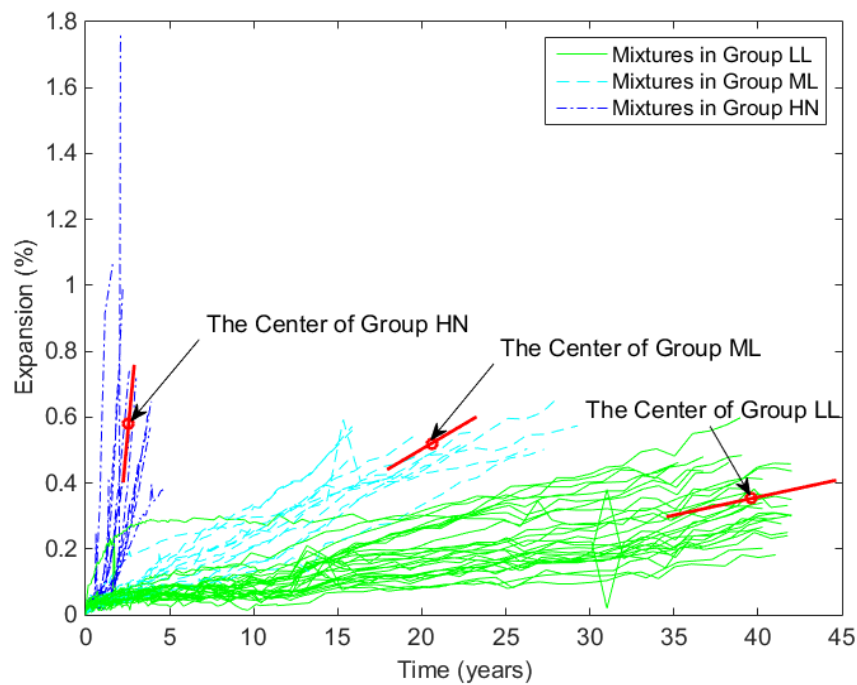

Fig. 1. The three groups of mixtures that were divided based on the expansion data using K-means clustering

**3.2 Suitable selection of variables for regression based on PCA**

There are seven main variables for each concrete mixture; accordingly, if all seven variables are considered, the regression will be excessively complicated. Therefore, it is necessary to determine which variables are the most important and will correspondingly affect the results the most. PCA is very helpful in this scenario. The PCA results are shown in Table 1 using the processed data of each group. Only the first three components of each group are selected because they contribute 84.8%, 94.5%, and 86.0% of the total variance in their corresponding groups, which means that those three components contain at least 85% of the data from their respective groups.

Table 1 The PCA results of three mixture groups

| Group | | HN | | | ML | | | LL | | |
|---|---|---|---|---|---|---|---|---|---|---|
| Sample size | | 12 | | | 14 | | | 28 | | |
| Order of principal components | | First | Second | Third | First | Second | Third | First | Second | Third |
| Coefficients of variables | W/C ratio | -0.460 | -0.135 | 0.234 | **0.581*** | -0.413 | -0.271 | -0.288 | **0.660*** | -0.169 |
| | Content of $C_3A$ | -0.001 | -0.109 | -0.526 | 0.342 | **0.679*** | 0.445 | **0.742*** | 0.072 | -0.019 |
| | Content of $C_3S$ | -0.004 | -0.023 | **0.5888** | -0.045 | 0.362 | -0.507 | -0.094 | 0.272 | **0.646*** |
| | Content of $C_2S$ | -0.128 | -0.069 | -0.557 | 0.052 | -0.380 | **0.517*** | -0.161 | -0.278 | -0.638 |
| | Content of $C_4AF$ | **0.602*** | 0.623 | 0.018 | 0.567 | -0.142 | -0.020 | 0.497 | -0.025 | 0.029 |
| | Volume of cement | -0.377 | **0.653*** | -0.110 | -0.457 | -0.251 | -0.025 | -0.094 | -0.574 | 0.307 |
| | Content of air | 0.517 | -0.386 | -0.026 | -0.100 | -0.097 | 0.450 | 0.275 | 0.278 | -0.228 |
| Contribution to total variance | | 94.5% | | | 86.0% | | | 84.8% | | |

According to the theory of PCA, the absolute value of the coefficient within a certain principal component is proportional to the ability of its corresponding variable to influence the result. Thus, we select the coefficient with the largest absolute value from each principal component (marked in Table 1) and consider its corresponding variable to be useful in the regression. In this way, three important variables are selected for each group. For Group HN, the content of $C_4AF$, the volume of cement and the content of $C_3S$ are the three principal variables; for Group ML, the W/C, content of $C_3A$ and content of $C_2S$ are the three principal variables; and for Group LL, the content of $C_3A$, W/C and content of $C_3S$ are the three principal variables. When comparing the results, it is quite striking that the components of Group ML and Group LL are completely different from those of Group HN, while those of the first two share some similarities with each other. This finding strongly indicates that Group HN has a completely different expansion mechanism from the other two groups; this might explain why the mixtures in Group HL expand nonlinearly while the other mixes exhibit linear curves. In addition, although Group ML shares two variables with Group LL, there are still some obvious differences between them; for example, these two groups have different third variables, and the first two variables are ordered in reverse. These differences prove that it is necessary to divide all the mixtures into three groups rather than two.

**3.3 Regression with the selected variables**

Before performing the regression, it is crucial to smooth the expansion curves of the mixtures in Groups ML and LL because even small errors will have substantial negative impacts on the curves due to the relatively small magnitudes. The convolution method is applied to smooth the curves according to the following equation:

$$X_n = \frac{T_{n+1,n}}{T_{n+1,n} + T_{n,n-1}}(1-\alpha)S_{n-1} + \alpha S_n + \frac{T_{n,n-1}}{T_{n+1,n} + T_{n,n-1}}(1-\alpha)S_{n+1} \quad (21)$$

where $X_n$ = processed value of point $n$

$S_n$ = original expansion value of point $n$

$T_{n+1,n}$ = interval of time between point $n$ and $n+1$

$\alpha$ = weight coefficient with a range from 0 to 1

$\alpha$ in the above equation balances the influence of the original value at a certain point with those of the values of its adjacent points on its processed value. After optimization, 0.3 is selected as the value of $\alpha$ in this research. Fig. 2 shows a comparison between the original and processed curves of mixture No. 1000.

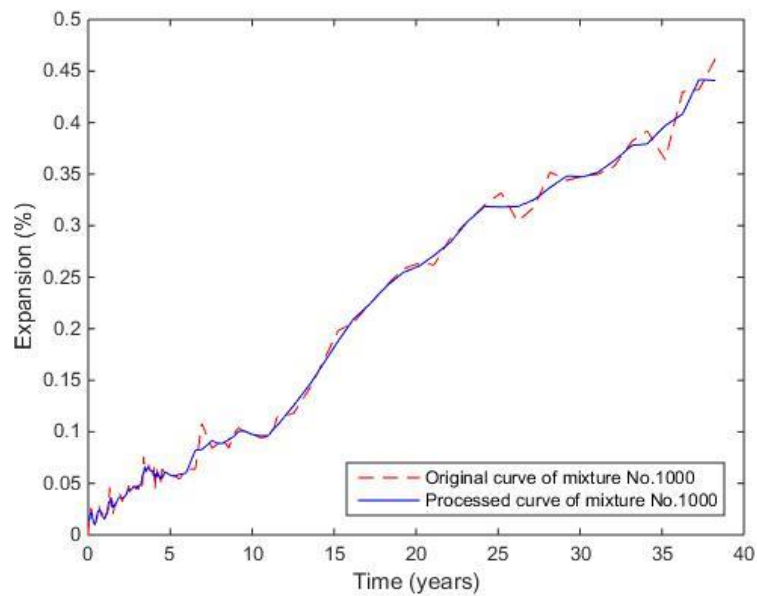

Fig. 2. The smoothed expansion curve of mixture No. 1000 compared with the original curve.

After the expansion curves have been preprocessed, linear regression is applied with the OLS method to obtain the regression functions of three groups of mixtures. Several combinations of potential variables obtained by the PCA algorithm are tested, and the optimal combination for each group is chosen.

For Group LL and Group LM, the two most important variables, namely, $C_3A$ and W/C, are chosen initially. However, the results of several regression attempts show that the influence of W/C in the regression is overwhelming, while the content of $C_3A$ introduces a difference of less than 5% in Group LL; however, the same results are not achieved in Group ML. This finding again proves the necessity of differentiating Group ML from Group LL. Therefore, the content of $C_3A$ is eliminated from the regression function of Group LL, while it is retained in Group ML. The variables are multiplied by time to be consistent with the time series data. Thus,

the model of Group ML takes the following form:
$$EXP = \alpha_1(WC * T) + \alpha_2(C_3A) * T + \alpha_3 \quad (22)$$
The model of Group LL is similarly expressed as follows:
$$EXP = \alpha_1(WC * T) + \alpha_3 \quad (23)$$
where
$$EXP = Expansion\ rate\ of\ the\ mix (from\ 0 - 100)$$
$$WC = Water\ to\ cement\ ratio (from\ 0 - 1)$$
$$C_3A = Percentage\ of\ C_3A (from\ 0 - 100)$$
$$T = Time\ of\ expasnion (year)$$

The parameter estimation results are shown in Table 2 (Group LL) and 3 (Group LM). The $R^2$ statistics of Group LL and LM are 0.7735 and 0.9197 (close to one), respectively, indicating that both regressions show a relatively high overall fit. When referring to a single variable, the conclusion that all explanatory variables included in the two models are statistically significant (i.e., not zero) is quite firm. The signs of all variables in both groups are positive, signifying that an increase in either the content of $C_3A$ (only in the Group ML model) or the W/C value leads to an increase in the expansion of concrete, and this expansion increases with time. Theoretically, this finding is correct because several studies regarding sulfate attack on concrete have reported that a high content of $C_3A$ is associated with the expansion of concrete.

The final model of Group LL is:
$$EXP = 0.0157(WC * T) + 0.0305 \quad (24)$$
The final model of Group ML is:
$$EXP = 0.0293(WC * T) + 0.000975(C_3A) * T + 0.0216 \quad (25)$$

**Table 2—Regression of Group LL**

| Dependent variable = expansion |  |  |
|---|---|---|
| Model size: observations = 2545; parameters = 2 | | |
| Residuals: standard deviation = 0.0021 | | |
| Fit: $R^2 = 0.7735$ | | |
| Variable | Coefficient | T-statistic |
| T*WC | 0.157E-01 | 92.086 |
| Constant | 0.305E-01 | 25.073 |

**Table 3—Regression of Group ML**

| Dependent variable = expansion |||
|---|---|---|
| Model size: observations = 870; parameters = 3 |||
| Residuals: standard deviation = 0.0015 |||
| Fit: $R^2 = 0.9197$ |||
| Variable | Coefficient | T-statistic |
| $T*WC$ | 0.293E-01 | 39.704 |
| $T*C_3A$ | 0.975E-03 | 14.427 |
| Constant | 0.216E-01 | 12.197 |

Meanwhile, for Group HN, the most important variable, namely, the content of cement, is selected as the only variable in the regression function because the cement content already fulfills the accuracy requirement in the model without other parameters according to trial calculations. Because of the nonlinear relationship between the expansion rate and time observed for the expansion curve of the mixtures in Group HN, the regression function is transformed into a nonlinear function by evaluating the logarithm of the expansion rate. In addition, another parameter consisting only of time is added to the function to cope with unknown nonlinear complexities. Thus, the form of the model for Group HN is as follows:

$$ln(EXP) = \alpha_1(CC * T) + \alpha_2 T + \alpha_3 \quad (26)$$

where
$EXP = Expansion\ rate\ of\ the\ mixture\ (from\ 0\ to\ 100)$
$CC = Content\ of\ cement\ (from\ 0\ to\ 1)$
$T = Time\ of\ expasnion\ (years)$

**Table 4—Regression of Group HN**

| Dependent variable = ln(expansion) |||
|---|---|---|
| Model size: observations = 517; parameters = 3 |||
| Residuals: standard deviation = 0.3550 |||
| Fit: $R^2 = 0.7256$ |||
| Variable | Coefficient | T-statistic |
| $T*CC$ | 11.20 | 7.345 |
| $T$ | -5.68 | -6.323 |
| Constant | -3.66 | -77.691 |

The results shown in Table 4 indicate a good fit (the $R^2$ value is 0.7256), even with the nonlinear model. Additionally, all of the parameters have a satisfactory T-statistic that is obviously different from zero. In this model, the content of cement affects the expansion that is similar to those of the W/C value and the content of $C_3A$ in the linear model, and thus, the cement content has a positive effect on the expansion of the concrete cylinders.

Therefore, the final model for Group HN is:
$$ln(EXP) = 11.20(CC * T) - 5.68T - 3.66 \quad (27)$$

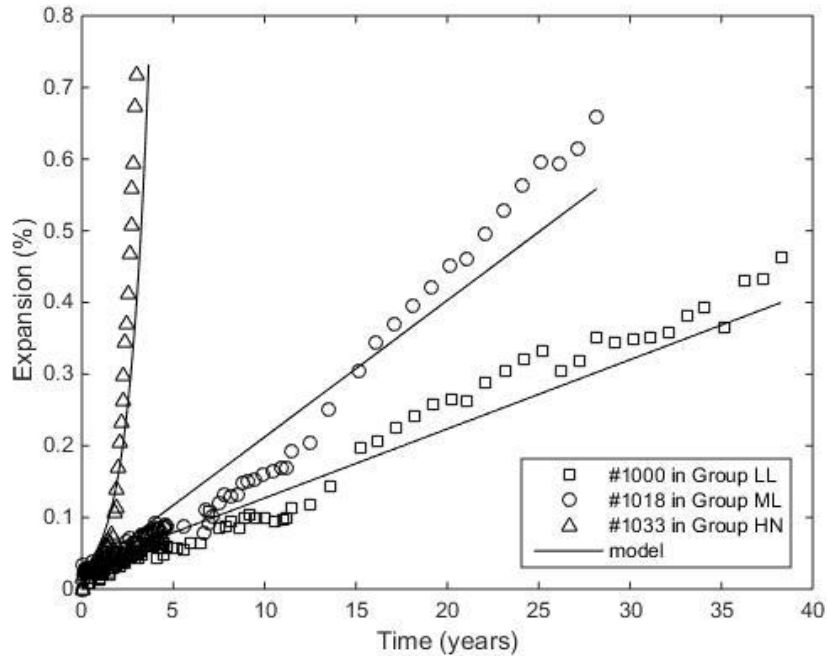

Fig. 3. Model predictions for three representative mixtures from the three groups and the original expansion data from the USBR specimens

Fig. 3 shows the predicted expansion percentages of three representative mixtures from Groups LL, ML and HN using equations (24), (25) and (27), respectively. Their experimental expansion data are also plotted in a scatter plot to facilitate comparison. The representative mixtures from among the USBR specimens are as follows: for Group LL, mixture No. 1000 (with a W/C value of 0.490); for Group ML, mixture No. 1018 (with a W/C value of 0.481 and a $C_3A$ content of 5.1%); and for Group HN, mixture No. 1033 (with a cement content of 0.589, i.e., greater than one).

As shown in Fig. 3, both the linear and the nonlinear predicted expansion curves fit the original data well, which strongly supports the conclusion that the model predictions accurately approximate the actual expansion behavior throughout the test period that exceeds 40 years.

**3.4 Reclassification boundaries**

Although the obtained regression functions have a satisfactory accuracy, the regression models still cannot be applied directly because there is no distinct criterion for classifying a certain mixture into one of the three groups mentioned above. The particular model of a certain group is valid only when dealing with a mixture belonging to that group. Therefore, before predicting the expansion curve of any mixture, that mixture must first be classified into one of the three groups. However, it is impossible to classify a mixture whose expansion curve still needs to be predicted. For the initial classification in this study, the K-means clustering method is based on the experimental performance of the mixture, which is obtained over a long-term 40-year testing period. Such a classification will increase the accuracy of the regression models but will not be available for a mixture without any experimental data whose expansion result we want to predict.

Thus, a new classification method that can classify a mixture based only on its material properties (e.g., W/C value and content of $C_3A$) must be developed; we call this process reclassification. To be more exact, the boundaries between the three groups will be purely dependent on the properties and parameters of the mixtures rather than the expansion data. With these boundaries, we will be able to classify every mixture in the database (i.e., the 54 specimens corresponding to the mixtures in this study) into a corresponding group, and the results will be consistent with the abovementioned results of the K-means clustering. Once we have the group boundaries, we will be able to classify a new mixture into one of three groups and find an adequate regression model with which to predict its expansion curve. To achieve this method, a linear SVM model, as mentioned above, is applied in this research to find a linear boundary that is an exact linear combination of the mixture properties and parameters.

According to relevant studies on time to failure and empirical models, the content of $C_3A$ and W/C have been regarded as the most important parameters that determine the expansion pattern of a concrete mixture. Therefore, in this research, these two parameters are also selected, which means that the boundary will be given in a 2-D plane with coordinates that consist of the $C_3A$ content (x-axis) and W/C (y-axis).

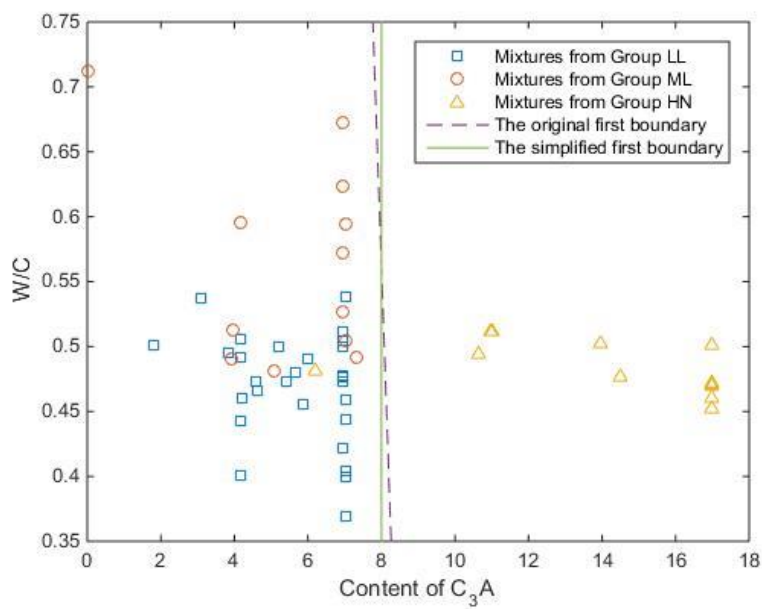

Fig. 4. The original and simplified boundaries separating the mixtures in Group HN from those in Groups LL and ML in the plane of $C_3A$ content and W/C

The coordinates of every mixture in the abovementioned plane are shown in Fig. 4, which demonstrates that the mixtures in Group HN are easily differentiated from those in the other two groups. As a result, the first boundary is established with the help of the SVM algorithm to separate Group HN from Groups LL and ML. Clearly, the mixtures in Group HN are not linearly separable from those in Groups LL and ML. Therefore, the box constraint C, which balances the error penalty with the overfitting tendency, is needed and is set as 100 in the model. Finally, the function of the first boundary is obtained as follows:

$$C_3A + 1.241 * WC - 8.697 = 0 \quad (28)$$

The function of the boundary line strongly indicates that it is almost parallel to the y-axis (i.e., W/C) when the magnitude of the range in W/C is compared with that in the content of $C_3A$. This result can also be proven by drawing the original boundary in Fig. 4. Thus, for the sake of convenience in engineering applications, the boundary is simplified as a line parallel to the y-axis. The simplified function of the boundary (also shown in Fig. 4 as the simplified first boundary) is:

$$C_3A = 8.00 \quad (29)$$

The meaning of the function is straightforward: a mixture with a content of $C_3A$ exceeding 8% will be classified into Group HN; otherwise, it will be classified into one of the other two groups.

After differentiating the mixtures in Group HN, we still need one more boundary, which we call the second boundary, to differentiate the mixtures in Group LL from those in Group ML.

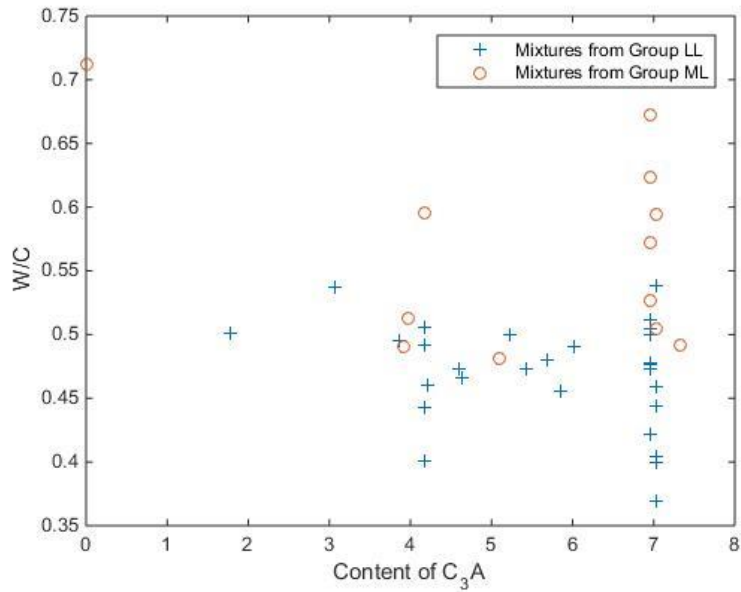

Fig. 5. Mixture distribution in the plane consisting of $C_3A$ content and W/C

As before, the remaining mixtures are drawn in Fig. 5, which demonstrates that the mixtures are still not linearly separable; moreover, the mixtures are interspersed too closely. Therefore, it will be immensely difficult to find a reasonable boundary in the plane consisting of $C_3A$ content and W/C. Accordingly, at least one axis must be changed. In this study, the content of $C_3S$ is chosen to replace the content of $C_3A$ as the new x-axis. The new plane and the new points that represent all of the mixtures are presented in Fig. 6, and the SVM algorithm is again applied with the box constraint C equal to 100. The form of the second boundary line function is:

$$C_3S + 387.3 * WC - 233.6 = 0 \quad (30)$$

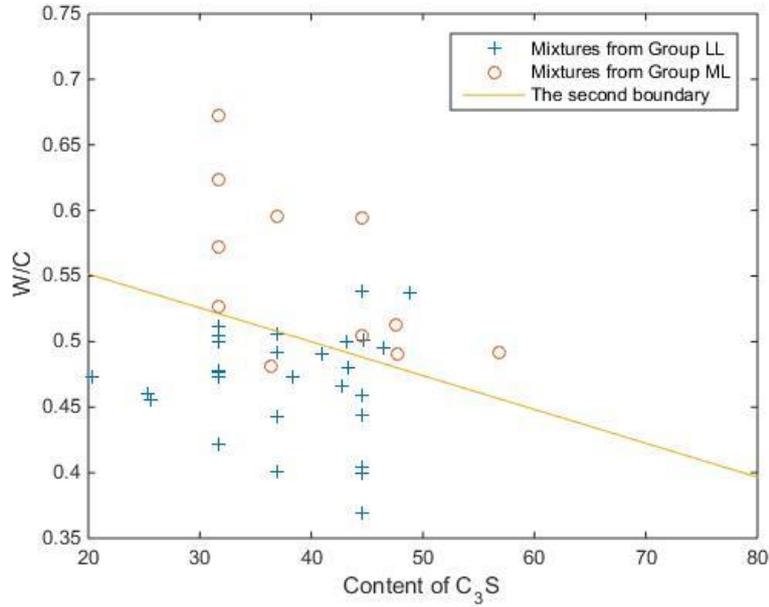

Fig. 6. The second boundary that separates the mixtures in Group LL from those in Group ML in the plane consisting of $C_3S$ content and W/C

A mixture with a corresponding point above the second boundary line will be placed into Group ML; otherwise, it will be classified into Group LL. However, similar to the situation with the first boundary, some mixtures are still being placed in groups that are different from their original group. This type of error cannot be avoided due to the limitations of our SVM algorithm; however, these errors will not substantially affect the accuracy of the regression model, as will be discussed later. To provide a clearer view of the classification results based on both the first boundary and the second boundary, the following summary is given:

$$A\ mix\ belongs\ to \begin{cases} Group\ HN, & if\ C_3A > 8\% \\ Group\ ML, & if\ C_3A > 8\%\ and\ C_3S + 387.3 * WC - 233.6 \geq 0 \\ Group\ LL, & if\ C_3A > 8\%\ and\ C_3S + 387.3 * WC - 233.6 < 0 \end{cases} \quad (31)$$

After finding both boundary lines, two issues still need to be considered. The first issue is that the accuracy of the regression function might be influenced because some mixtures, which we call error mixtures, have been classified into a group that is different from their original group. Thus, these error mixtures will be predicted by an erroneous regression function, i.e., a function for an incorrect group. However, this issue may not be important; based on the division of groups that is given by the SVM algorithm using the mixture properties instead of the K-means clustering results using expansion data, the $R^2$ statistics of the regression functions of the three groups still maintain high: 0.7460 (Group LL), 0.9076 (Group ML) and 0.7351 (Group HN). The maximum decrease in $R^2$ is less than 5%, and there is even an increase in the $R^2$ value of Group HN. The result shows that the proposed regression model exhibits strong generalizability.

The second issue is that a sensitivity analysis must be performed on the classification results obtained by the SVM algorithm to determine whether they rely excessively on the data used to train the model. To achieve this analysis, several redundant specimens with a certain mixture proportion that has already appeared in this study are used as data to verify the model. First, their expansion data are introduced into the K-means clustering algorithm to determine the group to which they belong. Next, we determine the group to which they belong according to the two reclassification boundaries. Then, based on whether consistent grouping results are observed among the results of the two classification methods, we call determine whether the proposed model is stable.

In total, 15 redundant specimens are considered in the analysis. Among them, 7 specimens belong to Group LL, 4 belong to Group ML and 4 belong to Group HN according to the K-means clustering results. The results of reclassification show that 12 of the 15 specimens are grouped consistently with a successful percentage of 80%, which is an acceptable number for the linear SVM model. Therefore, it is reasonable to consider that our model is not excessively sensitive to changes in the training data and has a satisfactory stability.

## 4 Conclusion

Models for predicting the expansion of concrete materials in severe, sulfate-rich environments were based on a panel dataset consisting of 6000 expansion measurements that were collected over a period exceeding 40 years by the USBR. The dataset included 69 cylindrical specimens produced from 54 mixture proportions. Through an analysis, all concrete mixtures are divided into three groups with distinct expansion patterns under sulfate attack. These mixtures are mainly different with regard to the time (in years) before failure, expansion speed and linearity of the expansion curve. A certain regression function is established for each group that fits the experimental expansion data of the mixtures in that group. During the regression process, PCA is employed to find the most significant variables and parameters in the regression function of each group. W/C and $C_3A$ content are the most important variables for mixtures with low and moderate expansion speeds, while the content of cement plays an important role in mixtures with a high expansion speed. After building the regression model, a criterion is developed using SVM, which is capable of classifying a mixture into one of three groups by applying a suitable regression function to the mixture. The classification criterion consists of two boundaries based only on the W/C, content of $C_3A$ and content of $C_3S$ of a mixture. Finally, an accuracy test and sensitivity analysis are performed on the model, the results of which show that the proposed model has satisfactory stability and generalizability.